\documentstyle[twocolumn,aps,epsfig]{revtex}

\begin{document}

\draft

\title{Laser Induced  Condensation  of Bosonic Gases in Traps}

\author{L. Santos$^{1}$, Z. Idziaszek$^{1,2}$,  
J. I. Cirac$^3$, and M. Lewenstein$^{1}$}

\address{(1) Institut f\"ur Theoretische Physik, Universit\"at Hannover,
 D-30167 Hannover,Germany\\
(2) Centrum Fizyki Teoretycznej, Polska Akademia Nauk, 02-668 Warsaw, Poland\\
(3) Institut f\"ur Theoretische Physik, Universit\"at Innsbruck,
 A--6020 Innsbruck, Austria}

\maketitle

\begin{abstract}
We consider collective laser cooling of atomic gas 
in the {\it Festina lente} regime, when the heating effects associated with 
photon reabsorptions are suppressed. We demonstrate that 
an appropriate sequence of laser pulses allows to
condense a gas of trapped bosonic atoms into the ground level of the
trap in the presence of collisions.
Such condensation is robust and can be achieved in
experimentally feasible traps. We extend significantly the validity 
of our previous numerical studies, 
and present new analytic results concerning condensation in 
the limit of rapid thermalization. We discuss in detail 
necessary conditions to realize all optical
 condensate in weak condensation 
regime and beyond. 
\end{abstract}

\pacs{32.80Pj, 42.50Vk}

%\narrowtext

\section{Introduction}

Laser cooling has led  to spectacular results in
recent years \cite{nobel}. So far, however, it has not allowed to reach
temperatures for which quantum statistics become important.
In particular, evaporative  cooling is used to obtain
Bose-Einstein condensation of trapped gases \cite{BEC}.
Nevertheless, several groups are pursuing the challenging goal of
condensation via all--optical means \cite{Salomon,Ertmer,Mlynek,new}.
The interest in laser--induced condensation stems from the fact that, 
(a) a laser--cooled system is open and is not necessarilly in thermal equilibrium, i.e. 
its physics is in principle 
 much richer than that of evaporatively--cooled gases; 
(b) laser cooling is perhaps the only method which may 
condense all atoms into the 
ground state, i.e. reach an effective zero temperature.

In traps of size larger than the inverse
wavevector, $k_L^{-1}$, the temperatures required for condensation are
typically below the photon recoil energy,
$E_R=\hbar\omega_R=\hbar^2 k_L^2/2m$, where $m$ is the atomic mass.
There exist several laser cooling schemes to reach such temperatures
cite{VSCPT,Raman}. They exploit  single atom
``dark states'', i.e. states which cannot be excited by the
laser, but can be populated via spontaneous emission. The main
difficulty in applying dark state cooling  for dense
gases is caused by light reabsorption. Indeed, these
states are not dark with respect to the photons spontaneously emitted by
other atoms. Thus, at sufficiently high densities, dark state cooling
may cease to work, since multiple reabsorptions can increase the system
energy by several recoil energies per atom
\cite{Sesko,Dalibard,Burnett,Ellinger}. In particular, laser induced
condensation is feasible only if the reabsorption probability is smaller
than the inverse of the number of energy levels accessible via
spontaneous emission processes \cite{Dalibard}.

Several remedies to the reabsorption problem have been
proposed. First, the role of reabsorptions increases with the
dimensionality. If the reabsorption cross section for trapped
atoms is the same as in free space, i.e. $\simeq 1/k_L^2$, the
reabsorptions should not cause any problem in 1D, have to be carefully
considered in 2D, and forbid condensation in 3D. Working with quasi-1D
or -2D systems is thus a possible way to reduce the role of
reabsorptions \cite{Janicke}.  The most promising remedy
against  this problem employs the
dependence of the reabsorption probability for trapped atoms on the
fluorescence rate $\gamma$, which can be easily adjusted in
dark state cooling \cite{reab}. In particular, in the so called {\em
Festina lente} limit, when $\gamma$ is much smaller than the trap
frequency $\omega$ \cite{Festina}, the reabsorption processes in which
the atoms change energy and undergo heating are suppressed.
This regime can be achieved either in Raman cooling \cite{Marzoli}
by changing the intensity and/or the detuning of the repumping laser,
or in systems which simply posses a narrow natural linewidth.

In a series of papers \cite{Morigi,1Atom,Manyatoms,Coll} we have 
investigated  collective cooling schemes in traps of realistic size
in the {\it Festina lente} limit, and we have  shown that
laser induced condensation is possible. In this paper  
we present also such investigation taking into account the
effects of atom--atom collisions. This paper is, in a sense, a final paper in 
a series of papers about  the {\it Festina lente} regime. In the first two 
of those papers we have generalized dark state (Raman) cooling method 
beyond the Lamb-Dicke regime, i.e. beyond the regime 
when the size of the trap 
is smaller than the laser wavelength\cite{Morigi,1Atom}. In particular, in
the second of these references we have demonstrated the possibility of 
cooling to an arbitrary state of the trap. We have then applied 
those methods to
samples of atoms and demonstrated possibility of all optical condensation
first for the case of an ideal gas \cite{Manyatoms}, and then taking 
into account elastic interatomic collisions\cite{Coll}.  
In all of 
the above paper we have limited, however, our investigations to {\it 
relatively 
small traps} with Lamb-Dicke parameter $\eta$ not greater  than 5 in 1D, and 
not greater than 2 in 3D because of numerical complexity. 
Also, we have considered so far only the {\it 
regime of weak condensation}, when the energy 
levels of the trapped gas are not yet modified by the 
mean field interactions. That implied that we could have considered 
only moderate numbers of atoms in the trap of the order of few hundred up to 
one thousand.

In this paper we overcome the above mentioned shortcomings. 
In particular:
\begin{itemize}

\item We extend our numerical studies to traps with $\eta=8$ 
in 1D and $\eta=4$ in 3D. 

\item We present analytic description of the 
dynamics in the weak condensation regime and in the 
limit of rapid thermalization due to elastic collisions. 

\item We extend our theory beyond the weak condensation 
regime to the hydrodynamic regime,  and present 
the analytic description of the dynamics in the limit of 
rapid thermalization using Bogoliubov-Hartree-Fock 
theory at finite temperature.

\item Finally, we present a detailed discussion of requirements
that have to be fulfilled in order to achieve experimentally 
all optical condensation in the {\it Festina lente} regime. 
 
\end{itemize}

The paper is organized as follows. The sections II--IV 
deal with the weak condensation regime. In section II, we formulate the
Master Equation (ME) that describes the combined effects of dark state 
(Raman) cooling 
in the Festina Lente regime, and elastic atom--atom collisions. Cooling 
is dynamical, and consists of
sequences of pairs of laser pulses inducing stimulated and
spontaneous Raman transitions between the two electronic levels
of trapped atoms, $|g\rangle$ and $|e\rangle$. The stimulated Raman 
transition induces the energy selective transition
$|g\rangle\to|e\rangle$ that depopulates all motional states except the
ground state which is dark; the spontaneous Raman transition via
a third auxiliary level, similarly as standard spontaneous emission, is 
non-selective, and repumps
the atoms from $|e\rangle$ to $|g\rangle$ populating all accessible
motional states. The collisions introduce a thermalization mechanism
which redistribute the populations of the trap levels according to a 
Bose--Einstein distribution (BED). In section III 
we present our numerical results. 
We simulate the dynamics generated by the ME in 1D and 3D, and
show that laser induced condensation into the 
ground state of the trap is not only possible with collisions, 
but it is even more robust when they are present. In Section IV 
we present analytic results concerning the limit of rapid thermalization, 
in which the state of the system at each instant 
can be described by the thermal 
density matrix of the canonical ensemble. 
We demonstrate here that laser cooling leads to final temperature of the
order of $E_R$ or less.

In section V we extend our theory beyond the weak condensation regime.
This can be done in the rapid thermalization limit, since in that case
one can describe the state of the system at each instant 
to be thermal and well 
described by the Bogoliubov--Hartree--Fock theory 
at finite temperature $T$ \cite{BHF}. Laser cooling introduces then a slow 
process of decrease of the effective temperature. 
We show that cooling down 
to $T\simeq 0$ is also possible beyond the weak condensation regime.
This allows us to present a detailed discussion of the prospects 
for all optical BEC in Section VI.

\section{Quantum Master Equation} 
 
We consider $N$ bosonic atoms with two levels
$|g\rangle$ and $|e\rangle$ in a non-isotropic dipole trap with the 
frequencies
$\omega^g_{x,y,z},\omega^e_{x,y,z}$ different for the ground and the
excited states, and non-commensurable one with another. This assumption
simplifies enormously the dynamics of the spontaneous emission processes
in the Festina Lente limit. We assume weak absorption pulses, so that
no significant excited--state population is present. This allows to
adiabatically eliminate the excited--state contribution, and
consequently to consider the density matrix
$\rho(t)$ describing all atoms in the ground state $|g\rangle$, and 
diagonal in the Fock representation corresponding to the bare trap levels.
The ground
state $|g\rangle$ into which the condensation will take place, should
be the ground electronic state of the atom, in such
a way that the inelastic collisions of two $|g\rangle$ atoms are not
possible \cite{Inellastic}. The Raman lasers are red--detuned
sufficiently far from molecular resonances, such that photoassociation
losses can be neglected for the regime of atomic densities considered,
$n<10^{15}$atoms/cm$^3$ \cite{Photo}. For these densities, three--body
losses are also negligible. We shall show that in the {\em weak
  condensation regime}, when the mean--field energy
$E_I<\hbar\omega$, the laser--cooling and collisional effects can
be considered separately. In particular, the
latter ones can be described by a
quantum Boltzmann ME \cite{gard,gard2}. 
The extension of our results beyond the weak condensation regime
 ($E_I>\hbar\omega$) is discussed in section V.

In the following we follow the same notation as in Refs.\ 
\cite{Manyatoms,Coll}. Let $g_{m}$, $g_{m}^{\dag}$ ($e_{l}$,
$e_{l}^{\dag}$) be the annihilation and creation
operators of atoms in the ground (excited) state and in the trap level
$m$ $(l) $, which fulfill the bosonic commutation relations. 
Using the standard theory of quantum stochastic processes 
one can develop the quantum ME which 
describes the atom dynamics in the Festina Lente regime \cite{Manyatoms}
\begin{equation}
\dot\rho(t)={\cal L}_0\rho+{\cal L}_1\rho+{\cal L}_2\rho,
 \label{ME}
\end{equation}
where
\begin{mathletters}
\begin{eqnarray}
{\cal L}_0\rho&=&-i\hat H_{eff}\rho(t)+i\rho(t)\hat H_{eff}^{\dag}+{\cal J}\rho(t), \\
{\cal L}_1\rho&=&-i[\hat H_{las},\rho(t)], \\
{\cal L}_2\rho&=&-i[\hat H_{coll},\rho(t)], 
\end{eqnarray}
\end{mathletters}
with
\begin{mathletters}
\begin{eqnarray}
\hat H_{eff}&=&\sum_{m}\hbar\omega_{m}^{g}g_{m}^{\dag}g_{m}+\sum_{l}\hbar(\omega_{l}^{e}
-\delta)e_{l}^{\dag}e_{l} \nonumber \\
&-&i\hbar\gamma\int d\phi d\theta sin\theta {\cal W}(\theta,\phi) \nonumber \\
&\times& \sum_{l,m}|\eta_{lm}(\vec k)|^{2}e_{l}^{\dag}g_{m}g_{m}^{\dag}e_{l}, \label{Heff}\\
\hat H_{las}&=&\frac{\hbar\Omega}{2}\sum_{l,m}\eta_{lm}(k_{L})e_{l}^{\dag}g_{m}+H.c.,\\
\hat H_{coll}&=&\frac{1}{2}\sum_{m_1,m_2,m_3,m_4} \!\!\!\! U_{m_1,m_2,m_3,m_4}
g_{m_4}^{\dag}g_{m_3}^{\dag}g_{m_2}g_{m_1},\\
{\cal J}\rho(t)&=&2\hbar\gamma\int d\phi d\theta \sin\theta{\cal W}(\theta,\phi) \nonumber \\
&\times& \sum_{l,m}[\eta_{lm}^{\ast}(\vec k)g_{m}^{\dag}e_{l}]\rho(t)[\eta_{lm}(\vec
k)e_{l}^{\dag}g_{m}].
\end{eqnarray}
\end{mathletters}
Here, $2\gamma$ is the single--atom spontaneous emission rate,
$\Omega$ is the  Rabi frequency 
 associated with the atom transition and the laser field, 
 $\eta_{lm}(k_{L})=\langle e,l|e^{i\vec k_{L}\cdot\vec r}|g,m\rangle$
 are the Franck--Condon
  factors,  ${\cal W}(\theta,\phi)$ is the fluorescence dipole
  pattern, $\omega _m^g$ 
 ($\omega_l^e$) are the energies of the ground (excited) harmonic trap
 level $m $ ($l$), and $\delta$ is 
 the laser detuning from the atomic transition \cite{dip}. 
In the regime we want
to study, only $s$--wave scattering is important, 
and then:
\begin{equation}
U_{m_1,m_2,m_3,m_4}=\frac{4\pi\hbar^2a}{m}\int_{R^3}d^3x
\psi_{m_4}^{\ast}\psi_{m_3}^{\ast}\psi_{m_2}\psi_{m_1},
\end{equation}
where $\psi_{m_j}$ denotes the harmonic oscillator wavefunctions, and $a$ is
the $s$-wave scattering length.

In the following we are going first to work in the weak--condensation
regime, where no mean--field effects are considered. 
In typical 
experiments this condition requires that the condensate cannot contain
more than $1000$ particles \cite{gard2}. 
We shall also consider that
$\Omega\ll\omega$. Due to 
the Festina--Lente requirements, $\Omega$ is in general smaller than
the typical collisional energy. Therefore we can formally establish 
the hierarchy ${\cal L}_0\gg{\cal L}_2>{\cal L}_1$. 

Let us project into the ground state configurations with $N_{j}$ atoms 
in the $j$--th level, and no excited atoms, $|\vec n\rangle\equiv
|N_{0},N_{1},\dots;g\rangle$$\otimes |0,0,\dots;e\rangle$, using the
projector operator ${\cal P}X=\sum_{\vec n}|\vec n\rangle\langle\vec
n|\langle\vec n|X|\vec n\rangle$. Using Born--Markov approximation as
in Ref.\ \cite{gard2}, and expanding up to order ${\cal
  O}({\cal L}_1^2)$, the ME for the dynamics of the reduced density operator 
$v={\cal P}\rho$ becomes:
\begin{equation}
\dot v(t)={\cal L}_{coll}v(t) + {\cal L}_{cool}v(t),
\label{ME2}
\end{equation}
where $
{\cal L}_{coll}=-{\cal PL}_2{\cal L}_0{\cal L}_2
$
describes the collisional part, and has the form of a QBME as that of
Refs.\ \cite{gard,gard2}. The laser--cooling dynamics is described by
\begin{eqnarray}
{\cal L}_{cool}&=&-{\cal PL}_{1}[{\cal L}_{0}]^{-1}{\cal L}_{1} \nonumber \\
&+&\frac{1}{2}{\cal PL}_0\int_{0}^{\infty}d\tau\!\int_{0}^{\infty}d\tau' e^{-{\cal
L}_{0}(\tau-\tau')}{\cal L}_{1}e^{-{\cal L}_{0}\tau'}{\cal L}_{1},
\label{Lcool}
\end{eqnarray}
which has the same form of the ME calculated for ideal--gas case \cite{Manyatoms}. 
The first correction to such splitting between both dynamics is of the order
${\cal L}_2{\cal L}_1^2[{\cal L}_{0}]^{-2}$, and  
therefore the independence between the collisions and laser cooling
dynamics is only valid in the weak--condensation regime. In the limit of fast 
thermalization beyond the weak condensation regime, we will employ the 
hierarchy ${\cal L}_0 + {\cal L}_2 \gg {\cal L}_1$.

\section{Numerical results}

Eq. (\ref{ME2}) can be simulated using standard
Monte Carlo procedures. We have first performed such simulations 
for a relatively small sample of $N=133$
atoms, and therefore well in the weak--condensation
regime. Franck-Condon factors and trap frequencies  can
be efficiently approximated using the states of an isotropic trap of
frequency $\omega$ \cite{1Atom}. We concentrate on 3D cooling, whose
analysis is
greatly simplified by considering ergodic approximation \cite{gard2},
i.e. considering that the
population of degenerate energy levels equalize on a time scale much
faster than the collisions between levels of different energies, and
than the laser--cooling time. The harmonic trap is considered beyond
the Lamb-Dicke
limit, i.e. the Lamb-Dicke parameter
$\eta=\sqrt{E_R/\hbar\omega}$ is larger than one.  Due to
memory storage and calculation times, our numerical simulation has to be
restricted to not too large values of $\eta$. In that case the atom
(having initially an energy of the order of few $E_R$) may increase
its trap energy level in the spontaneous emission process by energies
$\sim E_R$, and non-standard cooling schemes have to be used to avoid
such heating effects \cite{Morigi}. Generalization of the approach of
Ref. \cite{Morigi} allows to cool dynamically (i.e. by changing
absorption laser pulses appropriately) {\it individual atoms} to
arbitrary trap levels \cite{1Atom}.

Let us first consider the {\em ideal gas} case, i.e. $a=0$. This
approximation applies in principle only for very dilute atomic
samples. However,
it has been demonstrated recently that $a$ can be externally
modified by magnetic fields \cite{Ketterle}.
Theoretical proposals describing modifications of $a$ using 
 red--detuned lasers \cite{Fedichev}, or dc-electric
fields \cite{You} are also under experimental studies. 
These methods could lead to experimentally feasible 
situations in which $a\simeq 0$. In fact, very recent results of the
JILA group indicate the realization of an ``ideal'' gas of $^{85}$Rb
is indeed possible\cite{jila}.
In such a case, it has been shown
\cite{Manyatoms} that an appropriate dynamical cooling scheme would allow
to condense a collection of trapped bosons, not only into the ground
state of the trap, but also into excited ones. 

The
cooling process that we consider  consists of pulses 
of different frequencies, which are
specially designed to leave just one single state not emptied. Since
the level can be filled via spontaneous emission, the level acts as a
trapping state, where the condensation is induced. Particularly
important is that it is possible to design \cite{1Atom} two different
dark--state mechanisms based, respectively, on the properties of the
Frank--Condon factors, and on the
destructive interference between the absorption amplitudes of orthogonal
lasers ("interference"--dark--states). For a collection
of atoms, the bosonic enhancement allows for  faster cooling,
the dark--states become more robust with respect to changes 
of the physical parameters, and
the multistability phenomena may appear.

Using Monte Carlo methods, we have first 
analyzed the case of rather small trap, and studied 
the ground--state cooling of 133
atoms in a 3D isotropic harmonic trap with $\eta=2$, using 20 3D-energy
shells (i.e. 1771 trap levels). We consider the case of
$\gamma=0.04\omega$ and $\Omega=0.03\omega$ to assure 
the {\it Festina lente} conditions. This choice of parameters is rather 
restrictive. In practice, however, 
 much larger values of $\gamma$ and $\Omega$, of the order of, say, $\omega/2$
can be used; in fact the use of larger $\gamma$ and $\Omega$ is 
highly recommended in order to ``hurry up slowly'', i.e. avoid 
long cooling times and reabsorptions simultaneously. 

% Figure 1
\begin{figure}[ht]
\begin{center}\
\epsfxsize=4.0cm
\hspace{0mm}
\psfig{file=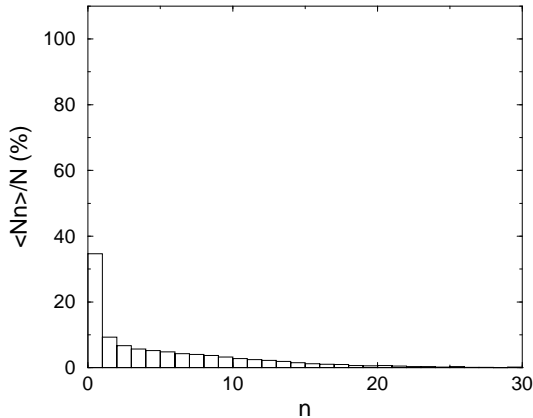,width=7.0cm}\\[0.1cm]
\caption{Average of the population of the different energy shells, after evolving the system 
under collisions, starting from a thermal distribution of $\langle n \rangle=6$.}
\label{fig:1}
\end{center}
\end{figure}

\begin{figure}[ht]
\begin{center}\
\epsfxsize=4.0cm
\hspace{0mm}
\psfig{file=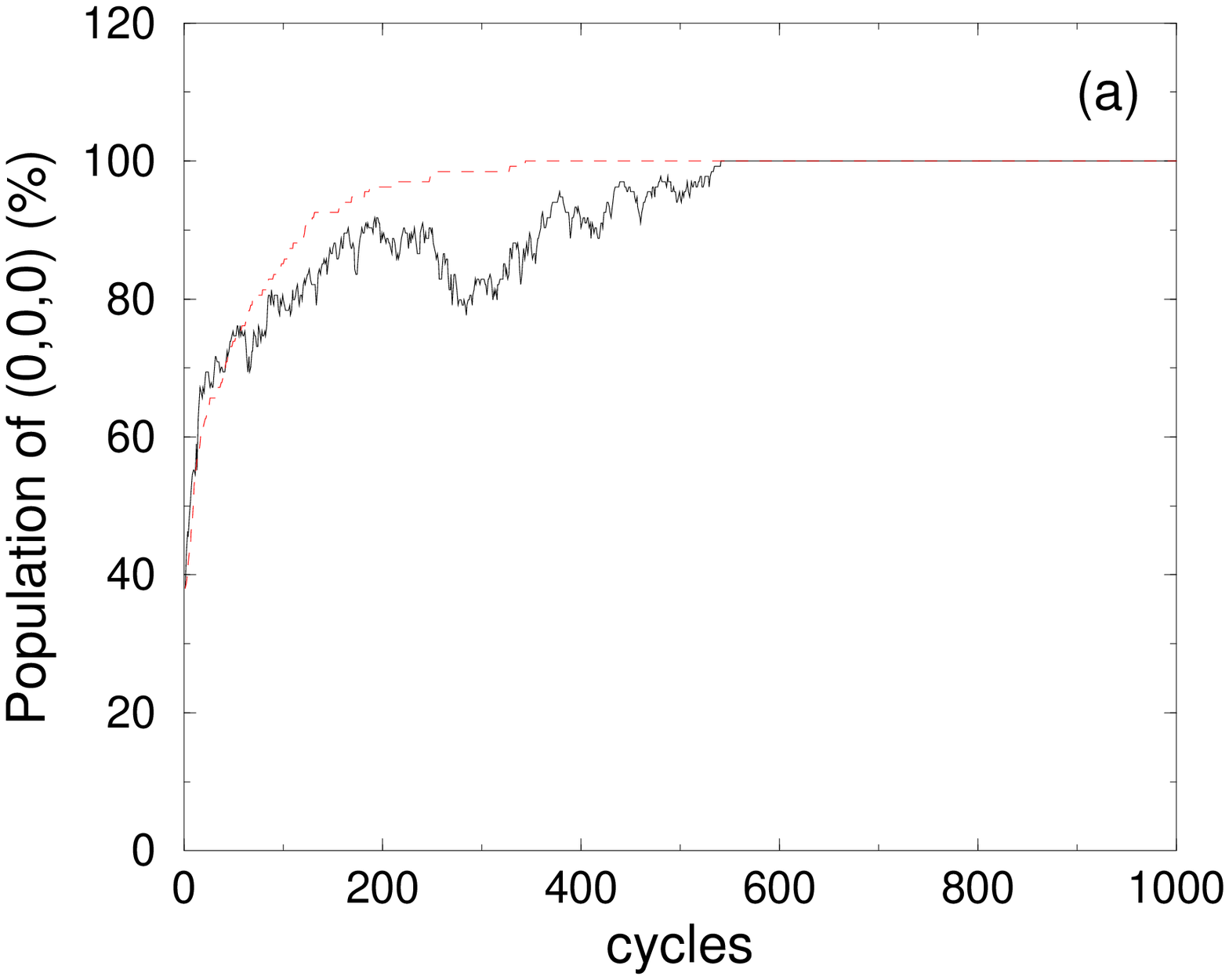,width=7.0cm}\\[0.1cm]
\psfig{file=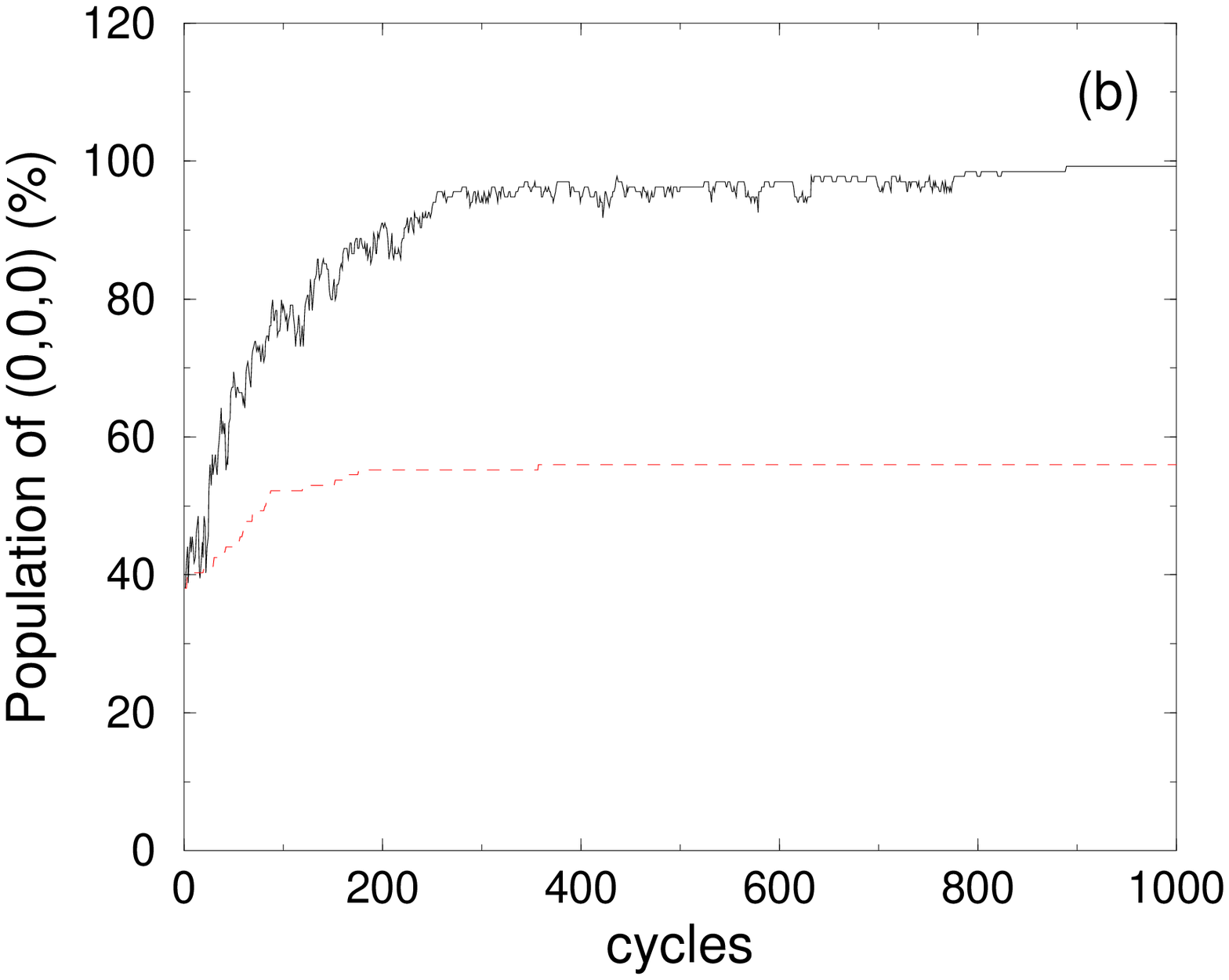,width=7.0cm}\\[0.1cm]
\caption{Population dynamics of the ground state of the trap for the case of
$\eta=2$, $\gamma=0.04\omega$, $\Omega=0.03\omega$, $N=133$, starting from the BED distribution of 
Fig.\ \ref{fig:1}.
(a) Cooling cycles composed of 
two laser pulses of detunings 
$\delta=s\omega$, with $s_{1,2}=-4,0$, and $A_z=1,-2$ 
respectively, are used. 
The non--ideal (solid line) and ideal (dashed
line) gas cases are compared. 
For $\omega = (2 \pi) \times 1$kHz the cycle duration is $\sim$ 0.028s. This time can be, 
however, significantly optimized as discussed in the text.
(b) Same case, but only employing pulse $1$.}
\label{fig:2}
\end{center}
\end{figure}

In order to compare our results with the case of finite 
$a>0$, we employ as an initial
atomic distribution the Bose--Einstein 
distribution (BED) (see Fig.\
\ref{fig:1}) obtained after evolving the system in presence of
collisions from an initial thermal distribution of mean
$6\hbar\omega$. As we see, it already contains quite substantial amount
of atoms condensed in the ground state, but also  
a lot of uncondensed ones. We must note that we begin with an BED 
below the condensation temperature due to numerical limitations, 
but qualitatively the same
results could be found if the initial BED would not have been condensed.
We shall demonstrate this point later.
The BED does not coincide exactly with the 
thermodynamical one described by grand canononical ensemble, 
due to finite--size effects.
Occupation numbers $N_n$ correspond to the ones obtained from the microcanonical
ensemble at finite $N$ and fixed energy. 
We apply our laser cooling cycles, each
composed by two laser 
pulses of detuning $\delta=s\omega$, with $s_{1,2}=-4,0$, and time duration 
$T=(2\gamma)/\Omega^2$. The laser pulses are emitted in three
orthogonal directions $x$, $y$ and $z$, and are 
characterized by their respective Rabi frequencies
$\Omega_x=\Omega_y=\Omega$, $\Omega_z=A_z\Omega$. For the first 
pulse we assume $A_z=1$, while for the second one $A_z=-2$ is
considered. With this choice, the second pulse is 
an ``interference'' dark--state pulse for the ground--state of the
trap. Fig.\ \ref{fig:2}(a) shows 
(dashed line) that these two 
pulses are able to condense the population into the ground state of
the trap, in absence of collisions; 
in particular no confinement pulses \cite{Morigi,Manyatoms,1Atom}(of
detuning $\delta=-3\eta^2\omega$) are needed. 
This is due to the 
bosonic enhancement, and the fact that initially the system is already
partially condensed.  The
dark--state pulse repumps the population in those states 
which are dark with respect to the pulses with detuning
$\delta=-4\omega$. During the dynamics the ground--state is the only
trap level which remains not emptied. As pointed out in Ref.\
\cite{1Atom}, the  many body effects introduce one very important
element to the dynamics: Bose enhancement factors, that speed up the
dynamics enormously.

% Figure 2

Let us consider now the case in which collisions
are taken into account. We shall consider the same situation and
laser--cooling scheme as previously. 
Fig.\ \ref{fig:2}(a) shows 
(solid line) the dynamics of the 
population of the ground-state in presence of collisions. After 600 cycles, 
all the population is transferred to the ground state of the 
trap. This means that applying the laser cooling scheme 
brings the system into an 
effective BED of $T=0$. It is easy to understand why 
the effect is maintained in presence of collisions, 
even considering that the collisional dynamics is much 
faster than laser--cooling.
The laser--cooling mechanism tends to 
decrease the energy per particle (i.e. the chemical potential of the system), 
in the same way as evaporative 
cooling does, but without the losses of particles in the trap during the process.
 Thermalization via 
collisions brings the system to a lower temperature. If one repeats the laser cooling
 sufficiently many times, the system 
ends with an effective zero temperature. Finally, let us point out that some auxiliary
 pulses which are needed 
in the ideal gas case, are not necessary in presence of collisions. In particular,
 for the previous
 example, the pulse of zero 
detuning (required for the ideal gas case, Fig.\ \ref{fig:2}(b) dashed line) 
is no more needed, 
as shown in Fig.\ \ref{fig:2}(b) (solid line). Thus,  
efficient laser--cooling is not only possible in presence of 
collisions, but it can even be significantly simplified.

We have been limited to rather small $\eta$ due to computational complexity.
In particular, in simulations discussed above 
$\eta=2$ in 3D is considered. 
The maximal value of $\eta$ we could reach in 3D recently was 4, 
and the corresponding results are 
presented in Fig. 3.

% Figure 3
\begin{figure}[ht]
\begin{center}\
\epsfxsize=4.0cm
\hspace{0mm}
\psfig{file=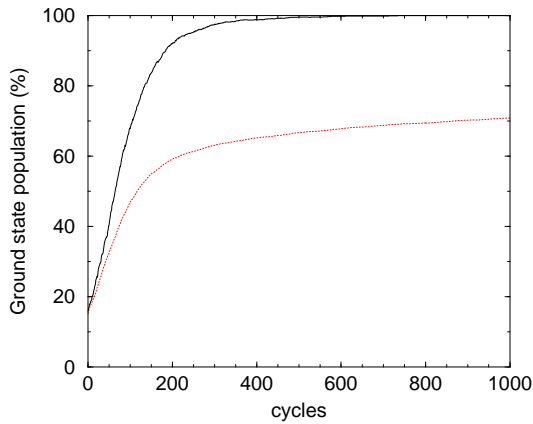,width=7.0cm}\\[0.1cm]
\caption{Population dynamics of the ground state of the trap for the case of
$\eta=4$, $\gamma=0.04\omega$, $\Omega=0.03\omega$, $N=500$, starting from the BED distribution 
with mean energy 12 $\hbar \omega$. Cooling cycles composed of 
two laser pulses of detunings $\delta=s\omega$, 
with $s_{1,2}=-16,0$, and $A_z=1,-2$ respectively, are used. The results for noninteracting 
(dotted line) and interacting (solid line) bosons are presented.
For $\omega = (2 \pi) \times 1$kHz the cycle duration is $\sim$ 0.028s. This time can be, 
however, significantly optimized as discussed in the text.
}
\label{fig:3}
\end{center}
\end{figure}

The Monte-Carlo simulations for $\eta=4$ have been performed for 500 atoms 
occupying 33 3D-energy shells (i. e. 6545 energy levels). 
The initial mean energy per atom is equal to 12 $\hbar \omega$, which is above the condensation point 
(mean energy corresponding
to $T_c$ is about $2.7 \hbar \omega$). The initial number of condensed atoms is, however, relatively large 
(about 10\%), what makes the cooling dynamics sufficiently fast from the beginning. 
The sequence of cooling pulses is similar as in the previous 3D calculations. 
The first pulse with detuning $\delta=-16 \omega$ and
$A_z = 1$ is responsible for confining the atoms, the second one with $\delta=0$ and $A_z = -2$ is the 
dark-state 
pulse. This combination of pulses is probably not the most efficient in the case 
of ideal gas, what may be noticed from Fig.\ \ref{fig:3}. After about 200 cycles, the populations of the 
first excited levels become significantly large. 
Those levels begin to compete with the ground state and the cooling efficiency is reduced. 
This effect 
does not occur in the presence of interactions. The collisions distribute the atoms in more 
uniform way among the excited states, and the condensation is approached quite fast.
     
\begin{figure}[ht]
\begin{center}\
\epsfxsize=4.0cm
\hspace{0mm}
\psfig{file=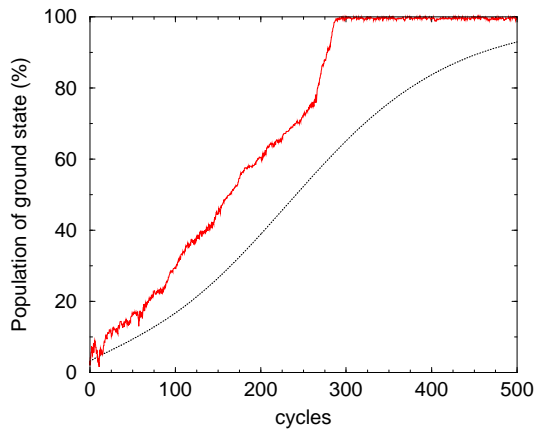,width=7.0cm}\\[0.1cm]
\caption{Population dynamics of the ground state of the 1D trap for 
the case of
$\eta=8$, $\gamma=0.5\omega$, $\Omega=0.4\omega$, 
$N=500$, starting from the BED distribution 
with mean energy $50\hbar \omega$. Cooling cycles composed of 
four laser pulses of detunings $\delta=s\omega$, 
with $s_{1,2,3,4}=-64,-65,37,39$ 
respectively, are used. The interacting case (solid line) and noninteracting case
(dotted line) are compared.
For $\omega = (2 \pi) \times 1$kHz the cycle duration is $\sim$ 0.056s. This time can be, 
however, significantly optimized as discussed in the text
}
\label{fig:4}
\end{center}
\end{figure}

2D and 1D simulations show that same  results hold 
for larger $\eta$. In particular, in 1D we have been able to study the 
cases up to $\eta=8$, which is illustrated in Fig.\ \ref{fig:4}.
For the presented case, the values of parameters are chosen   
close to the limits of the applicability of our approach. This choice of the parameters 
leads to shorter cooling times, keeping Festina Lente condition.
The pulse length is the same as in the previous simulations, in order to simplify 
the comparison of the cooling times.  
The presence of interactions accelerate the cooling process, similarly to the considered 
3D examples.
The studies for the different values of Lamb-Dicke parameter 
allowed us to observe 
that our results scale with $\eta$.
In 3D, for instance,  keeping $N$ fixed  the density 
scales as $\eta^{-3}$, the three--body loss rate as
$\eta^{-6}$,  $\omega$  
as $\eta^{-2}$, and the cooling time as $\eta^{2}$ times the number of necessary cycles.
The estimation of the latter is discussed in Sec. VI. 

% Figure 4

\section{Rapid thermalization in the weak condensation regime}

From the above discussion it is clear that physically it is to be expected 
that collisions will act on a faster time scale then the laser cooling. 
In that case, collisions will provide a fast thermalization mechanism. 
The dynamics of the system will reduce to a fast approach toward instantaneous
thermal equilibrium, and a slow change of effective temperature $T$.
In order to describe these changes quantitatively, we have to write 
down explicitly kinetic equations describing populations of the trap levels. 
This is done, as mentioned above, by performing adiabatic elimination of 
the excited state, and is discussed in more details below. 

In the regime of parameters we 
consider, the most of the atoms are in the ground state during
the dynamics. In particular the system is affected by two different 
time scales, a slow one given
by the commutator with $\hat H_{las}$  and other given by the rest
of terms in the master equation. Due to this fact we can use adiabatic 
elimination techniques to remove the
excited-state populations. Let $|\vec n\rangle\equiv |N_{0},N_{1},\dots;g\rangle$$\otimes
|0,0,\dots;e\rangle$ the ground state configurations with $N_{j}$ atoms in the $j$th level, and no
excited atoms; let their corresponding energies are $E_{\vec n}$,
 and their corresponding populations
$\rho_{\vec n\vec n}\equiv\langle\vec n|\rho |\vec n\rangle$. By using standard Projection Operator
techniques it is possible to show that the adiabatic elimination of the excited state levels (see
Appendix B of the  Ref. \cite{Manyatoms}) 
leads to a set of rate equations for the populations:
\begin{equation}
\dot\rho_{\vec n\vec n}(t)=\sum_{\vec m}P_{\vec n\leftarrow\vec m}\rho_{\vec m\vec m}(t)-\sum_{\vec
m}P_{\vec m\leftarrow\vec n}\rho_{\vec n\vec n}(t)+\dot\rho^{coll}_{\vec n\vec n}(t),
\label{rateeq1}
\end{equation}
where $P_{\vec m\leftarrow\vec n}$ are also defined in the Appendix of the Ref.\cite{Manyatoms},
 and can be understood
as the probabilities to undergo a transition 
from a ground--state configuration: $|\vec
n\rangle=|N_{0},N_{1},\dots\rangle$ to other configuration
$|\vec m\rangle$=$|N'_{0},N'_{1},\dots\rangle$. 

The last term in Eq. (\ref{rateeq1}) 
denotes the collisional contribution of the changes of $\rho_{\vec n\vec n}(t)$. 
A detailed discussion of the dynamics introduced by this term can be found in  
\cite{gard,gard2}. In our case, we do not have to 
specify it here: we just have to remember that in the regime of rapid thermalization, this terms
leads to a very fast thermalization of the system.

 Eq.\ (\ref{rateeq1}) can be rewritten in
terms of rate equations with respect to the populations $N_{j}$ in each level of the ground--state trap:
\begin{equation}
\dot N_{n}=\sum_{m}\Gamma_{n\leftarrow m}N_{m}-\sum_{m}\Gamma_{m\leftarrow n}N_{n}+\dot N_{n}^{coll},
\label{rateeq2}
\end{equation}
where the last term on RHS describes the contribution of collisions, while the 
rates are of the form:
\begin{eqnarray}
&&\Gamma_{n\leftarrow m}=\frac{\Omega^{2}}{2\gamma}\int_{0}^{2\pi}d\phi\int_{0}^{\pi}d\theta{\cal
W}(\theta,\phi) \nonumber \\
&&\times \left |\sum_{l}\frac{\gamma\eta_{ln}^{\ast}(\vec
k)\eta_{lm}(k_{L})}{[\delta-\omega(l-m)]+i\gamma R_{ml}}\right 
|^{2}(N_{n}+1-\delta_{n,m}).
\label{Gnm}
\end{eqnarray}
We have approximated here the excited and ground potentials 
by an harmonic potential of frequency
$\omega$, whereas:
\begin{eqnarray}
R_{ml}&=&\int_{0}^{2\pi}d\phi\int_{0}^{\pi}d\theta{\cal W}(\theta,\phi) 
\nonumber \\
&\times& \sum_{n'}|\eta_{ln'}(\vec k)|^{2}(N_{n'}+1-\delta_{n',m}).
\label{Rml}
\end{eqnarray}
To obtain the rates $\Gamma_{n\leftarrow m}$ we have used the 
properties of the creation and
annihilation operators when applied on Fock states. Note that Eq.\ (\ref{rateeq2}) is the same as
that we have found in \cite{1Atom} for the single--atom case. However, the
 rates (\ref{Gnm}) and the  single--atom
ones are quite different. For the single--atom case they coincide as they should, but for the
many--atom case, transition rates (\ref{Gnm}) become nonlinear, due to their dependence 
on the number of atoms in each trap level. In particular, we clearly 
distinguish here the two different quantum--statistical
contributions:
\begin{itemize}
\item In the denominator of the rates, we observe that the spontaneous emission acquires a
collective character (similar as that observed in superradiance). 
\item In the numerator of the rates, the 
 bosonic--en\-han\-ce\-ment factor $(N_{n}+1-\delta_{n,m})$ appears.
\end{itemize}
The bosonic--enhancement factor favors the atoms into just one level of the trap. 
The reason is that if we are able to pump a
significant amount of atoms into one single level $|n\rangle$, 
subsequent transitions into
$|n\rangle$ are more and more probable. Therefore the condensation is more effective.

The contribution of the collective spontaneous emission is unfortunately not so advantageous. The
cooling methods designed for the single--atom case, which we want to apply in the many--atom case,
are based on resonant processes, and are therefore quite dependent on a narrow resonance. Note that
the width of the resonance centered at $\delta=\omega(l-m)$ is given by $\gamma R_{ml}$. If
$\gamma R_{ml}$ grows, then the resonance is broadened, and the dark--state effects could
cease to exist. On the other hand, as $R_{ml}$ grows, the height of the
resonance becomes lower, and consequently the cooling becomes slower.
These negative effects can be avoided by using a sufficiently small $\gamma$, which is possible in
Raman cooling scheme. In Ref. \cite{Manyatoms} for instance,  
 the value 
$\gamma=0.005\omega$ was used in all the calculations, which guaranteed for one--dimensional
calculations that only the resonant terms were relevant. 
It has also been demonstrated that
the negative effects are less important for higher dimensions. Note that the
decreasing of $\gamma$ makes the cooling slower. In examples of $R_{ml}$'s discussed 
in \cite{Manyatoms}
it has led  to large cooling times (larger than $20$ seconds for sodium
atoms). This technical problem can be eventually solved by decreasing 
externally $\gamma$ during the
cooling process, in such a way that the maximum effective collective spontaneous emission rate
remains approximately constant and comparable to, say,  $\omega/2$. 
Such technical optimization would increase the cooling rate,
allowing for realistic cooling times, as we will show in Section IV.

\begin{figure}[ht]
\begin{center}\
\epsfxsize=4.0cm
\hspace{0mm}
\psfig{file=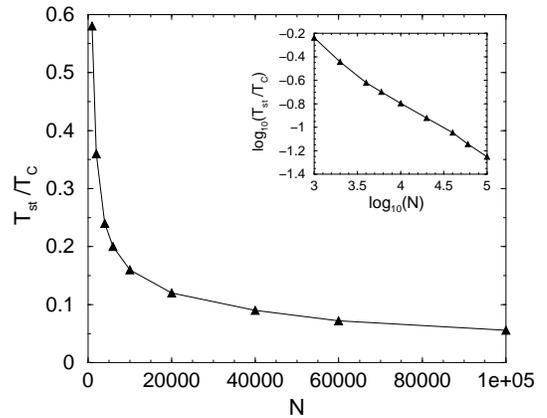,width=7.0cm}\\[0.1cm]
\caption{Stationary value of the temperature $T_{st}/T_c$ as a function
of $N$ for the case of $\eta=4$, $\gamma=0.04\omega$, $\Omega=0.03\omega$. 
The same results in logarithmic scale are presented in the inset.
}
\label{fig:5}
\end{center}
\end{figure}

In the limit of fast thermalization, we can assume that all $N_n$ adjust 
their values to an instantaneous thermal equilibrium 
characterized by the temperature $T(t)$,
i.e. are given by the Bose-Einstein distribution. 
The temperature varies slowly in time as the
laser cooling proceeds. Particularly  interesting
is the situation below critical temperature, in which we have
\begin{equation}
N_n=\exp(-\beta(t)\hbar\omega_n)/(1-\exp(-\beta(t)\hbar\omega_n)),
\label{eny}
\end{equation}
for $n \neq 0$, and
$$N_0=N(1-(T(t)/T_c)^3),$$
 where $\beta(t)=1/k_BT(t)$. The critical temperature is given by 
$$k_BT_c=\hbar\omega (N/g_3(1))^{1/3},$$
where the Bose function $g_3(1)\simeq 1.2$.
From the rate equations (\ref{rateeq2}) 
we immediately get the equation characterizing the slow 
changes of the bare energy of the system $E=
\sum_n \hbar\omega_n N_n$. 
Below the condensation point, $E(T)=3k_BT(k_BT/\hbar\omega)^3g_4(1)$, 
where $g_4(1)\simeq 1.08$, so that the above 
 Eqs. \ref{rateeq2} gives us the desired equation describing the changes 
of the temperature,
\begin{equation}
\dot E=\sum_n\hbar\omega_n\left[\sum_{m}\Gamma_{n\leftarrow m}N_{m}-\sum_{m}
\Gamma_{m\leftarrow n}N_{n}\right].
\label{rateeq3}
\end{equation}
We can easily rewrite this equation as
\begin{equation}
\frac{dT}{dt}=F(T),
\end{equation}
where
\begin{eqnarray}
F(T)&=& \frac{T}{4E(T)}\nonumber \\
&\times&\sum_n\hbar\omega_n\left[\sum_{m}\Gamma_{n\leftarrow m}N_{m}
-\sum_{m} \Gamma_{m\leftarrow n}N_{n}\right].
\label{rateeq3f}
\end{eqnarray}
Assuming that the ground state is perfectly dark, i.e. $\Gamma_{m\leftarrow 0}
=0$ for all $m$, 
it is easy to see that this equation has a (stable) stationary solution $T=0$. 
If the ground state is not perfectly  dark, 
one obtains a finite stationary value $T_{st}$. 
The stationary temperature is plotted against $N$ in Fig.\ \ref{fig:5}.
If $N$ is sufficiently large $T_{st}$ is very small, as we expect.
 The cooling rate 
during approach to $T \simeq 0$ is of the order of typical $\Gamma_{0\leftarrow m}$, 
and becomes significantly slower only at very low temperatures for very 
large $N$ (compare the discussion in \cite{Manyatoms}).  

% Figure 5

The value of $T_{st}$ can be estimated assuming that 
the final distribution of $N_n$ will more or less 
balance the laser cooling dynamics. Similarly as in Ref. \cite{Manyatoms},
we can consider the limit when $N_0\simeq N$, $N_m\ll N_0$ for 
$m\ne 0$ in a self-consistent way. In that limit the expected 
stationary values of $N_m$ are
$$ N_n\simeq  \Gamma_{n\leftarrow 0}N_0/\Gamma_{0\leftarrow n}. $$
We have $\Gamma_{n\leftarrow 0}\sim O(1)$, $\Gamma_{0\leftarrow n}\sim O(N_0)$,
so that $N_n\sim O(\gamma/\omega)$ 
in the stationary state. In this limit, the dynamics consists of 
independent jumps $0\to n\to 0$ \cite{Manyatoms}, and only the band 
of $n$'s 
corresponding to roughly two recoil energies is 
relevant ($n\le 2 E_R/\hbar\omega
=2 \eta^2$). 
From the expression
$$ \sum_{m\ne 0}^{m\le 2 \eta^2}N_m=N-N_0, $$
we estimate then
$$ \frac{T_{st}}{T_c}=\left(\sum_{m\ne 0}^{m\le 2 \eta^2}N_m/N\right)^{1/3}, $$
which implies that $k_BT_{st}\simeq k_BT_c\eta^2(\gamma/\omega N)^{1/3}\simeq
(\gamma/\omega)^{1/3}E_R$. As we see as $N\to \infty$, $T_{st}/T_c\to 0$ as 
$1/N^{1/3}$. 
This estimation provides in fact only an upper bound for $T_{st}$, 
due to the fact that it does not take into account Franck-Condon coefficients.
Numerical calculations 
for $\eta=4$, $\gamma=0.04\omega$ and $\Omega=0.03\omega$ presented in Fig. 5 clearly indicates 
then $T_{st} \rightarrow 0$ even faster as 
$1/ \sqrt{N}$ for $N \rightarrow \infty$.
We shall show in the next section that similar
 conclusions hold beyond the weak condensation.

\section{Beyond the weak condensation regime}

Let us first point out that beyond the weak condensation regime, 
the mean--field energy cannot be
neglected. This has two--fold 
consequences: (i) The levels of the trap are non--harmonic, i.e. they are not equally 
separated, because their energies become dependent on 
the occupation numbers; (ii) 
the wavefunctions are different, and in particular the condensate wavefunction 
becomes broader (we consider here only $a>0$). The fact 
that the energy levels 
are not harmonic any more, complicates the 
laser cooling, but the use of pulses 
with a variable frequency and band--width should produce the same  results 
as those presented here. 
The point (ii) implies that the central density of the interacting gas 
is much lower than the one predicted for noninteracting particles, and
therefore the dangerous regime of $n\simeq 5\times10^{14}$ atoms/cm$^3$ 
is reached for larger
number of atoms than in the ideal case. For example, for the case of
Magnesium, wavelength $\lambda=600$nm and $\eta=8$, the mentioned density regime 
is reached for just
$N=3.5\times 10^3$, whereas in Thomas--Fermi approximation the same is
true for $N=6.6\times 10^5$. Below this number, 
the interaction between the particles is dominated by 
the elastic two--body collisions 
considered in this paper. We show below that if our laser 
cooling scheme could be extended 
beyond the weak--condensation regime, laser--induced 
condensation of more than $10^6$ atoms would be feasible. 

In general, to extend the quantum master kinetic theory beyond the weak condensation 
theory is a formidable task (compare \cite{gard,gard2}). In general we do not know how the 
many body states of the trapped gas change in the course of the dynamics. 
The situation simplifies, however, in the limit of fast thermalization. 
The collisions assure then that the instantaneous 
thermal equilibrium is achieved within a fast time scale. Such 
equilibrium can be very 
well described for a weakly interacting Bose gas using the 
Bogoliubov-Hartree-Fock (BHF) theory \cite{BHF}. 

\subsection{Bogoliubov-Hartree-Fock theory}

The BHF theory has various versions; for our purposes perhaps the best approach is the 
one valid for a fixed number of atoms developed by Gardiner\cite{gphase}, and Castin and Dum
\cite{cphase}. In those approaches one does not break the phase symmetry, and avoid thus 
the time dependent effects associated with the spreading of the phase distribution \cite{mlyou}.
Unfortunately, the number conserving approach is technically tedious. It is therefore easier 
to use the standard BHF theory, with a broken phase symmetry and in the Popov approximation.
In this case we have to neglect the effects of slow spreading of the phase distribution.
This approach will be used below.
In the standard BHF theory at finite $T$, we introduce the quasi--particle annihilation and creation operators,
which are related via the unitary Bogoliubov transform to the atomic quantum field operators \cite{mlyou}:
\begin{eqnarray}
\tilde g_k&=&\int d^3x \left[u^*_k(x)\hat\Psi(x)+v_k(x)\Psi^{\dag}(x)\right],\\
\tilde g^{\dag}_k&=&\int d^3x \left[u_k(x)\hat\Psi^{\dag}(x)+v^*_k(x)
\Psi(x)\right],
\label{bogo}
\end{eqnarray}
where $u_k$ and $v_k$ fulfill the Bogoliubov--de Gennes equations with the eigenenergy 
$\hbar\tilde\omega_k$. In particular, $g_0$, $g_0^{\dag}$ are   associated with annihilation 
or creation of condensed particles.  
Using the bare states of the trap, the above relation can be rewritten as
\begin{eqnarray}
\tilde g_k&=&\sum_n \left[u^*_{kn}g_n+v_{kn}g^{\dag}_n\right],\\
\tilde g^{\dag}_k&=&\sum_n \left[u_{kn}g^{\dag}_n+v^*_{kn}g_n\right],
\label{bogo1}
\end{eqnarray} 
where $u_{kn}=\int d^3x \ u^*_k(x) \psi_n(x)$. The above equations can be inverted to obtain
\begin{eqnarray}
g_n&=&\sum_k \left[u_{kn}\tilde g_k-v_{kn}\tilde g^{\dag}_k\right],\\
 g^{\dag}_n&=&\sum_k \left[u^*_{kn}\tilde g^{\dag}_k
-v^*_{kn}\tilde g_k\right].
\label{bogo2}
\end{eqnarray} 

The instantaneous equilibrium corresponds to a 
thermal state of the system described by the 
approximated Hamiltonian $\sum_k\hbar\tilde\omega_k \tilde g_k^{\dag}\tilde g_k$. In order 
to describe the cooling dynamics we have to express the quantum master equation of section
II in terms of the BHF operators. We need to do it only for the terms responsible for laser cooling,
since the collisional and free parts are responsible only for reaching the instantaneous equilibrium. 
Following the same steps as in previous section, we perform the 
adiabatic elimination of the
excited state, carefully replacing particles by quasiparticles at each step.

\subsection{Rate equations for populations}

 Let $|\vec n\rangle\equiv |N_{0},N_{1},\dots;g\rangle$$\otimes
|0,0,\dots;e\rangle$ be the ground state configurations with $N_{n}$ quasiparticles 
in the $n$th level, and no
excited atoms; let their corresponding energies are $\tilde E_{\vec n}$,
 and their corresponding populations
$\rho_{\vec n\vec n}\equiv\langle\vec n|\rho |\vec n\rangle$. By using the similar 
Projection Operator
techniques as in the previous section
 it is possible to show that the adiabatic elimination of the excited state levels  
leads to a set of rate equations for the populations:
\begin{equation}
\dot\rho_{\vec n\vec n}(t)=\sum_{\vec m}P_{\vec n\leftarrow\vec m}
\rho_{\vec m\vec m}(t)-\sum_{\vec
m}P_{\vec m\leftarrow\vec n}\rho_{\vec n\vec n}(t)+
\dot\rho^{coll}_{\vec n\vec n}(t),
\label{rateeq1b}
\end{equation}
where $P_{\vec m\leftarrow\vec n}$ have similar form 
as those that appear in the weak condensation limit
\begin{eqnarray}
&&P_{\vec n\leftarrow\vec m}=2\gamma\int d\phi d\theta sin\theta {\cal
W}(\theta,\phi)\sum_{s,l} \nonumber \\
&&\times \left |\langle\vec n|(\tilde\eta_{ls}^{\ast}(\vec
k)\tilde g_{s}^{\dag}+ \tilde\zeta_{ls}(\vec k)\tilde g_s)e_{l}
\frac{1}{E_{\vec m}-\hat H_{eff}}\hat H_{las}^{(+)}
|\vec m\rangle \right|^{2},
\label{rate2b}
\end{eqnarray}
where $\tilde \eta_{ls}(\vec k)=\int d^3x \psi^*_l(x)e^{i\vec k\cdot\vec x}
u_s(x)=\sum_{s'}\eta_{ls'}u_{ss'}$, 
and $\tilde \zeta_{ls}(\vec k)=\int d^3x \psi^*_l(x)e^{i\vec k\cdot\vec x}
v_s(x)=\sum_{s'}\eta_{ls'}v_{ss'}$. The laser Hamiltonian is also 
reexpressed as 
$$
\hat H_{las}=\frac{\hbar\Omega}{2}\sum_{l,m}e_{l}^{\dag}
(\tilde\eta_{lm}(k_{L})g_{m}+\tilde\zeta^*_{lm}(k_{L})g^{\dag}_{m}).$$
As before, the probabilities (\ref{rate2b}) can be understood
as the probabilities to undergo a transition from a 
ground--state quasi-particle configuration: $|\vec
n\rangle=|N_{0},N_{1},\dots\rangle$ to other quasi--particle 
configuration
$|\vec m\rangle$=$|N'_{0},N'_{1},\dots\rangle$. 

The last term in Eq. (\ref{rateeq1b}) 
denotes the collisional contribution to 
the changes of $\rho_{\vec n\vec n}(t)$. We do not have to 
specify it here: as in Section IV we just have to remember that 
in the regime considered, this terms
leads to a very fast thermalization of the system.

 Eq.\ (\ref{rateeq1b}) can be rewritten in
terms of rate equations with respect to the quasi--particle populations 
$N_{n}$ in each level of the ground--state trap:
\begin{equation}
\dot N_{n}=\sum_{m}\Gamma_{n\leftarrow m}N_{m}-
\sum_{m}\Gamma_{m\leftarrow n}N_{n}+\dot N_{n}^{coll},
\label{rateeq2b}
\end{equation}
where the last term on RHS describes the contribution of 
collisions, while the 
rates are of the form:
\begin{eqnarray}
&&\Gamma_{n\leftarrow m}=\frac{\Omega^{2}}{2\gamma}\int_{0}^{2\pi}d\phi\int_{0}^{\pi}d\theta{\cal
W}(\theta,\phi) \nonumber \\
&&\times \left\{ \left |\sum_{l}\frac{\gamma\tilde\eta_{ln}^{\ast}(\vec
k)\tilde \eta_{lm}(k_{L})}{\delta-\omega^e_l+\tilde\omega_m+i(\gamma R_{ml}\!+\!\gamma_L)} \right|^{2}
\!(N_{n}+1-\delta_{n,m}) \right. \nonumber\\
&&+ \left. \left| \sum_{l}\frac{\gamma\tilde\zeta_{ln}(\vec
k)\tilde \zeta^*_{lm}(k_{L})}{\delta-\omega^e_l-\tilde\omega_m+i(\gamma R_{ml}\!+\!\gamma_L)}\right|^{2}
\!(N_{n}-\delta_{n,m}) \!\right\}
\label{Gnmb}
\end{eqnarray}
In the above expression we have used rotating wave approximation with respect to quasiparticle 
frequency spacings. One should stress that, although in general splittings
between the quasiparticle
energies  can be small, in the most interesting limit of large $N$, i.e. 
in the hydrodynamic regime the quasiparticle spectra are very regular,
 and the splittings are of the order of the bare  harmonic 
oscillator frequency $\omega$\cite{hydro}. That is why in the hydrodynamic regime the use of RWA with 
respect to the quasiparticle energy splittings is 
very well grounded. The widths in Eq. (\ref{Gnmb}) become now
\begin{eqnarray}
R_{ml}&=&\int_{0}^{2\pi}d\phi\int_{0}^{\pi}d\theta{\cal W}(\theta,\phi) 
\nonumber \\
&\times& \sum_{n'}|\tilde\eta_{ln'}(\vec k)|^{2}(N_{n'}+1-\delta_{n',m})\nonumber\\
&\times& \sum_{n'}|\tilde\zeta_{ln'}(\vec k)|^{2}(N_{n'}-\delta_{n',m}).
\label{Rmlb}
\end{eqnarray}
We have additionally included in the expression (\ref{Gnmb}) the width $\gamma_L$ which 
mimics the effects of the laser bandwidth, and which is
necessary to assure approximate fulfilling of the resonance 
conditions. The widths $\gamma_L$ should in practice be of the order 
of $\omega$. 

The equations derived in this section have a very 
similar form to the one derived in the previous sections. 
 In particular, we expect that in the Thomas-Fermi (hydrodynamic) regime, they will lead to 
the self--consistent equation for the effective temperature. 
We can easily rewrite this equation as
\begin{equation}
\frac{dT}{dt}=F(T),
\end{equation}
where
\begin{eqnarray}
F(T)&=&\frac{1}{E^\prime(T)} \nonumber \\
&\times& \sum_n\hbar\tilde\omega_n\left[\sum_{m}\Gamma_{n\leftarrow m}N_{m}
-\sum_{m} \Gamma_{m\leftarrow n}N_{n}\right] ,
\label{rateeq3b}
\end{eqnarray}
where $N_m$ are distributed according to the Bose--Einstein distribution 
corresponding to the quasiparticle Hamiltonian, whereas
$E(T)=\sum_n \hbar\tilde\omega_n N_n$, $E^\prime(T)=dE/dT$.
 Note that temperature 
dependence enters the above equations explicitly, and through the 
dependence of the quasiparticles and their energies on the temperature in the 
BHF theory at finite $T$. In the limit of large $N$, the same 
argumentation as in previous section can be applied.  We expect 
that the system will be cooled down to the temperature $T_{st}\simeq 0$, 
and that the stationary temperature $T_{st}$ will 
be of the order of $k_BT_{st}\simeq k_BT_c\eta^2(\gamma/\omega N)^{1/3}\simeq
(\gamma/\omega)^{1/3}E_R$. As we see as $N\to \infty$, $T_{st}/T_c\to 0$ at
least as fast as $1/N^{1/3}$. 

\section{Prospects for all optical BEC}

According to our results the prospects for all optical BEC are very good provided
several precautions are realized in experiments. The prescription to achieve all optical BEC is:

\begin{itemize}

\item Use a dipole trap of the moderate size $\eta \simeq 2-8$ and $\omega \sim 2 \pi
\times 1$ kHz.

\item Trap and cool atoms into the electronic hyperfine ground state. This allows to eliminate 
non-elastic two body collisions from considerations.

\item Work in the Festina-Lente limit to avoid reabsorptions. Use either natural narrow 
line, or quenched (Raman) transition of width $\gamma_{eff} < \omega$. In fact 
to shorten the cooling dynamics try to work with $\gamma_{eff} \sim \omega/2$.

\item work with red detuned laser tuned below or in between the molecular resonances 
to avoid photoassociation losses.

\item Avoid high densities, i. e. limit the number of trapped atoms $N$ in such a way that 
$n < 5 \cdot 10^{14}/{\rm cm}^3$. This will allow you to avoid 3-body losses and remaining 
photoassociation losses. 

\end{itemize}

In the Tab.\ \ref{tab:1} we present our estimates for the maximal number of condensed atoms 
and the cooling time. 
We have considered here Mg atoms, and assumed $\lambda_L=600$nm and scattering length 
$a_{sc} \simeq 5$ nm. For various values of $\eta$ we have then calculated corresponding trap 
frequency $\omega$. Maximal number of atoms $N_{max}$ has been estimated from requirement
that the corresponding peak density should be $5 \cdot 10^{14}/{\rm cm}^3$. The calculations has been 
done for the ideal ($a_{sc} = 0$) and weakly interacting gas, where for simplicity 
the Thomas-Fermi density profile was assumed. 

For the estimation of the cooling time $t_{cool}$ we use the following reasoning. The probability 
of atom jump (transfer) from level $m$ to level $n$ is proportional to the occupation numbers:
$p_{m\rightarrow n}\sim N_m (N_n + 1)$. If we omit the dependence on Frank-Condon coefficients, the 
jump probability is given by $p_{m \rightarrow n} =  {\cal N} N_m (N_n + 1)$, where ${\cal N}$ should be 
determined from the normalization condition: $\sum_m \sum_n p_{m\rightarrow n} = p_{tot}$, where
$p_{tot}$ is total probability of jump. Below we shall assume that in each cycle roughly 10\%
of atoms undergo a jump, i. e. $p_{tot}\simeq 0.1$.
We consider the final stage where only the dark-state pulse for the ground state is responsible for 
cooling. We assume that the ground state is perfectly dark, i. e. $p_{0\rightarrow m} = 0$
for all $m \neq 0$.
The spontaneous emission process scatters
the atoms within the energy band of width about $4 \eta^2 \hbar \omega = 4 E_R$.
Assuming that the dynamics takes place within the region of trap levels with energies 
smaller than  $4 \eta^2 \hbar \omega$, the normalization coefficient reads
\begin{equation}
{\cal N} = \frac{p_{tot}}{(N-N_0)(N+N_{st})} 
\label{Norma}
\end{equation}
where $N_{st}$ is the number of states with energies between 0 and $4 \eta^2 \hbar \omega$: 
$N_{st} \approx (4\eta^2)^3/6$. The cooling efficiency grows if we assume that in each cycle
 10\% of the excited states population is transferred. This number is consistent with the 
conditions for the adiabatic elimination used in our derivations. The population 
of the ground state is increased in every step by $(N-N_0)\sum_{m \neq 0} p_{m\rightarrow 0}$.
Thus we are able to write the formula describing the ground state dynamics:
\begin{equation}    
N_0(k+1)=N_0(k)+\epsilon(N-N_0(k))(N_0(k)+1),
\label{RecurencN0}
\end{equation}
where $\epsilon = (10(N+N_{st}))^{-1}$ and $k$ labels the cycles. 
In rough approximation the equation (\ref{RecurencN0}) may be replaced by the differential equation
\begin{equation}    
\frac{dN_0}{dk}=\epsilon (N-N_0)(N_0+1).
\label{DifferN0}
\end{equation}
Starting with 1 atom in the ground state, the number of cycles 
needed to obtain $N/2$ condensed atoms may be calculated to be 
$\ln(N)/(\epsilon N)$. Finally assuming  that each cycle have the duration $2/\gamma$, where
$\gamma \simeq \omega/4$, we estimate $t_{cool}\simeq 8 \ln(N)/(\omega \epsilon N)$. 
The analytic estimates agree quite well with numerical simulations of Sec. III. 
Direct inspection of the Tab.\ \ref{tab:1}, clearly shows that all-optical BEC of quite large 
number of atoms should be feasible within a reasonable cooling time.     

\section{Acknowledgments}

We acknowledge  support Deutsche Forschungsgemeinschaft (SFB
407), ESF PESC Proposal BEC2000+, and TMR ERBXTCT96-0002.
 We thank I. Bouchoule, W. Ertmer, E. Rasel, T. Mehlst\"aubler, J. Keupp,  K. Sengstock,  and Ch. Salomon
for fruitful discussions. We specially acknowledge discussions, help 
and valuable suggestions from Y. Castin, who took part in the earlier phase of
this project. 

\begin{table}[ht]
\begin{tabular}{rrrrr}

& \multicolumn{2}{c}{$a_{sc}=0$} & \multicolumn{2}{c}{$a_{sc}=5$nm} \\

\vspace{0.1cm} $\eta$ & \multicolumn{1}{c}{$N_{max}$} & 
\multicolumn{1}{c}{$t_{cool}$} & \multicolumn{1}{c}{$N_{max}$} 
& \multicolumn{1}{c}{$t_{cool}$} \\ \hline

2 & 55  & $119$ms  & 161  & $59.3$ms  \\
4 & 439  & $5.45$s  & $1.0 \times 10^{4}$  & $441$ms  \\
6 & $1.4 \times 10^3$  & $51.86$s  & $1.2 \times 10^5$  & $1.21$s  \\
8 & $3.5 \times 10^3$  & $232.4$s  & $6.6 \times 10^5$  & $2.5$s  

\end{tabular}
\vspace{0.6cm}

\caption{Maximal number of condensed atoms and cooling times calculated for Mg atoms, 
assuming $\lambda_L=600$nm and scattering length $a_{sc}\simeq 5$ nm}
\label{tab:1}
\end{table}


\begin{references}
\bibitem{Inellastic} For alkalis, one should apply a static magnetic
  field to split the levels in the lowest hyperfine manyfold;
  $|g\rangle$ and $|e\rangle$ are the lowest two Zeeman sublevels. For
  earth alkalis, $|g\rangle$ must be the ground state, and $|e\rangle$
  the excited with the ultranarrow (natural, or laser induced) linewidth.
\bibitem{dip} Consevative dipole--dipole interactions do not play a 
significant role at the considered densities ($<10^{15}/$cm$^2$), 
and are omitted in Eq.\ (\ref{Heff}). Their role is discussed in detail 
in Refs. \cite{Manyatoms,Coll}.
\end{references}
\end{document}